\documentclass[twocolumn,english,aps,prl,showpacs]{revtex4}
\usepackage[T1]{fontenc}
\usepackage[latin1]{inputenc}
\usepackage{amsmath}
\usepackage{amssymb}

\makeatletter

\usepackage[dvips]{graphicx}

\input{epsf}

\usepackage{babel}

\usepackage{babel}

\usepackage{babel}

\usepackage{babel}
\makeatother
\begin{document}

\title{Mean field phase diagrams of imbalanced Fermi gases near a Feshbach resonance}

\author{Hui Hu$^{1,2}$ and Xia-Ji Liu$^{2}$} 

\affiliation{$^{1}$\ Department of Physics, Renmin University of China, Beijing 100872, China \\
$^{2}$\ ARC Centre of Excellence for Quantum-Atom Optics, Department
of Physics, University of Queensland, Brisbane, Queensland 4072, Australia}

\date{\today{}}

\begin{abstract}
We propose phase diagrams for an imbalanced (unequal number of atoms or
Fermi surface in two pairing hyperfine states) gas of atomic fermions near a 
broad Feshbach resonance using mean field theory. Particularly, in the plane 
of interaction and polarization we determine the region for a phase separation phase 
composed of normal and superfluid components. We compare our prediction of
phase boundaries with the recent measurement, and find a good qualitative agreement.
\end{abstract}


\maketitle

Two recent experimental studies of fermionic superfluidity in strongly
interacting atomic $^6$Li gases with controlled population imbalance in two
spin components have attracted intense interest from the physicists in wide 
communities \cite{ketterle,hulet}. A very salient reason is the
mysterious nature of the pairing mechanism \cite{fflo,sarma,yip,bedaque,son,sheehy,carlson,rmp,he}.
Since the Bardeen-Cooper-Schrieffer (BCS) pairing requires an equal number
of atoms in each spin state, the presence of spin population imbalance
leads to some exotic forms of pairing, such as the finite-momentum paired
Fulde-Ferrell-Larkin-Ovchinnikov (FFLO) state \cite{fflo}, the
breached pairing or Sarma superfluidity \cite{sarma,yip}, and phase
separation \cite{bedaque}. However, the true ground state of imbalanced fermionic
superfluidity remains elusive and has been the subject of debate for
decades. The two recent experimental observations open up the intriguing
possibilities for resolving this long-standing problem. As the population 
imbalance increases, the disappearance of superfluidity has been 
identified \cite{ketterle}, and phase separation of a unitary gas in trap 
has been observed \cite{hulet}.

Motivated by the significant experimental development, in this paper 
we present a general mean field analysis of the ground state
of homogeneous imbalanced atomic gases, focusing on the strongly interacting
region near the broad Feshbach resonance, namely, the so-called crossover 
from BCS superfluidty to the Bose-Einstein condensation (BEC). Our goal 
is to map out the \emph{qualitative} zero-temperature phase
diagrams in the entire BCS-BEC crossover. A previous discussion of such
phase diagrams is based on a purely educated guess \cite{son}. Further
analytic mean-field estimate is restricted to the narrow Feshbach resonance 
\cite{sheehy}, for which the most fascinating crossover region has been
essentially ruled out, and thus is of less experimental relevance.

In contrast to these prior theoretical studies, our analysis is in close
connection to the experiment and has more predictive powers. Our main
results may be summarized as follows: \textbf{(1)} Aside from the ability to
include the exotic phases mentioned earlier, our mean-field calculation
predicts a new phase (the saddle point solution below), which becomes
energetically favorable for a finite population imbalance.
However, the new solution is inherently unstable towards phase separation,
signifying an inhomogeneous mixed phase. Around the crossover, in consistent 
with the experimental observations \cite{ketterle,hulet} we find that 
the phase separation phase becomes dominant in the phase diagram. 
\textbf{(2)} We construct the phase boundary of superfluid-to-normal transitions, 
and compare it with the measurement by Zwierlein \textit{et al.} \cite{ketterle}. 
The agreement is qualitatively good.

We consider an imbalanced Fermi gas of $^6$Li atoms across a broad Feshbach resonance, 
which is well described by using a single-channel model \cite{xiajipra}, 
\begin{equation}
{\cal H}=\sum_{\mathbf{k}\sigma }\xi _{\mathbf{k}\sigma }c_{\mathbf{k}\sigma }^{+}c_{%
\mathbf{k}\sigma }+g\sum_{\mathbf{kk}^{\prime }\mathbf{p}}c_{\mathbf{k}\uparrow }^{+}c_{%
\mathbf{p}-\mathbf{k}\downarrow }^{+}c_{\mathbf{p}-\mathbf{k}^{\prime }\downarrow }c_{%
\mathbf{k}^{\prime }\uparrow }.  \label{hami}
\end{equation}
Here the pseudospins $\sigma=\uparrow, \downarrow$ denote the two
hyperfine states of $^6$Li, and $c_{\mathbf{k}\sigma }^{+}$ is the fermionic
creation operator with the kinetic energy $\xi _{\mathbf{k}\sigma }=\epsilon _{%
\mathbf{k}}-\mu _\sigma $ and $\epsilon _{\mathbf{k}}=\hbar ^2\mathbf{k}^2/2m$. 
The chemical potentials are different, \textit{i.e.}, $\mu _{\uparrow ,\downarrow }=\mu \pm
\delta \mu $, to account for the population imbalance $\delta n=n_{\uparrow}-n_{\downarrow }$. 
$g$ is the bare interaction strength, and is expressed in terms of $s$-wave scattering length 
$a$ via $(4\pi \hbar ^2a/m)^{-1}=g^{-1}+$ $\sum_{\mathbf{k}}(2\epsilon _{\mathbf{k}})^{-1}$.

In the mean field approximation we decouple the interaction term by introducing
an order parameter of Cooper pairs in momentum space $\Delta =-g\sum_{\mathbf{k}}\left\langle c_{%
\mathbf{q/2}-\mathbf{k}\downarrow }c_{\mathbf{q/2}+\mathbf{k}\uparrow }\right\rangle $,
where the pairs may possess a nonzero center-of-mass momentum $\mathbf{q}$ in
case of spatially modulated states \cite{footnote}. As a result, the order parameter 
in real space aquires a one-wave oscillation form: 
$\Delta(\mathbf{x}) = -g \left\langle c_{\downarrow}(\mathbf{x})c_{\uparrow}(\mathbf{x}) \right\rangle = \Delta e^{i \mathbf{q} \cdot \mathbf{x}}$.
The value of $q$, together with $\Delta$, are to be determined. The Hamiltonian can then be
approximated by, 
\begin{eqnarray}
{\cal H} &=&\sum_{\mathbf{k}\sigma }\xi _{\mathbf{k}\sigma }c_{\mathbf{k}\sigma
}^{+}c_{\mathbf{k}\sigma }-\Delta \sum_{\mathbf{k}}\left[ c_{\frac{\mathbf{q}}{2%
}-\mathbf{k}\downarrow }c_{\frac{\mathbf{q}}{2}+\mathbf{k}\uparrow
}+h.c.\right] -\frac{\Delta ^2}g,  \nonumber \\
&=&\sum_{\mathbf{k}}\psi _{\mathbf{k}}^{+}\left[ \xi _{\mathbf{k}+}\sigma _z+\xi _{%
\mathbf{k}-}-\Delta \sigma _x\right] \psi _{\mathbf{k}}+{\cal E}_0,
\label{pairhami}
\end{eqnarray}
where in the second line we define a Nambu creation field operator: 
$\psi _{\mathbf{k}}^{+}=(c_{\mathbf{q/2}+\mathbf{k}\uparrow }^{+},c_{\mathbf{q/2}-\mathbf{k}\downarrow })$, 
$\sigma _x$ and $\sigma _z$ are the $2\times 2$ Pauli matrices, 
$\xi _{\mathbf{k}\pm }=(\xi _{\mathbf{q}/2+\mathbf{k}\uparrow }\pm \xi _{%
\mathbf{q}/2-\mathbf{k}\downarrow })/2$, and ${\cal E}_0=\sum_{\mathbf{k}}\left( \xi
_{\mathbf{k}+}-\xi _{\mathbf{k}-}\right) -\Delta ^2/g$. The above pairing
Hamiltonian may be solved by the standard Bogoliubov transformation, or more
straightforwardly, by employing the Nambu propagator $\mathbf{G}(\mathbf{k}%
,i\omega _m)=1/[(i\omega _m-\xi _{\mathbf{k}-})-\xi _{\mathbf{k}+}\sigma _z+\Delta
\sigma _x]$ with quasiparticle energies $E_{\mathbf{k}\pm }=(\xi _{\mathbf{k}%
+}^2+\Delta ^2)^{1/2}\pm \xi _{\mathbf{k}-}$. Here $\omega _m=(2m+1)\pi /\beta $ and $\beta =1/k_BT$. 
The thermodynamic potential thus takes the form, 
\begin{eqnarray}
\Omega &=&\frac 1\beta \sum_{\mathbf{k}m}\text{Tr}\ln \mathbf{G}(\mathbf{k},i\omega
_m)+{\cal E}_0,  \nonumber \\
&=&-\frac{m\Delta ^2}{4\pi \hbar ^2a}+\sum_{\mathbf{k}}\left[ \xi _{\mathbf{k}%
+}-\left( \xi _{\mathbf{k}+}^2+\Delta ^2\right) ^{1/2}+\frac{\Delta ^2}{%
2\epsilon _{\mathbf{k}}}\right]  \nonumber \\
&&+\frac 1\beta \sum_{\mathbf{k}}\left[ \ln f\left( -E_{\mathbf{k}+}\right) + \ln f\left(
-E_{\mathbf{k}-}\right) \right] ,  \label{omf}
\end{eqnarray}
where $f\left( x\right) =[\exp (\beta x)+1]^{-1}$ is the Fermi distribution
function. We shall confine ourselves to zero temperature, where the
last term in $\Omega$ reduces to $\sum_{\mathbf{k}}[E_{\mathbf{k}%
+}\Theta (-E_{\mathbf{k}+})+E_{\mathbf{k}-}\Theta (-E_{\mathbf{k}-})]$.

The mean field treatment presented above provides a simplest unified
description for the uniform and spatially modulated superfluids.
All these phases have to be determined using the stationary (\emph{saddle point}) 
conditions: $\partial \Omega /\partial \Delta =0$, $\partial \Omega/\partial q=0$, 
as well as the requirement of number conservation 
$n=n_{\uparrow }+n_{\downarrow }=-\partial \Omega /\partial \mu $.

\begin{figure}
\begin{center}\includegraphics*[width=7.0cm]{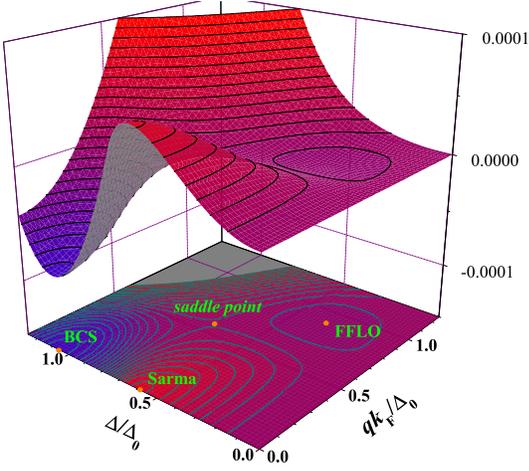}\end{center}

\caption{(color online). Landscape of the thermodynamic potential at $1/k_Fa=-1$. 
The chemical potential is fixed to $\mu =0.98942\epsilon _F$. The competing ground 
states are (i) a normal Fermi gas with $\Delta =0$, (ii) a fully paired BCS 
superfluid with $\Delta =\Delta _0$, $q=0$, and $\delta n=0$, (iii) a breached 
pairing or Sarma superfluid with $\Delta <\Delta _0$, $q=0$, and $\delta n\neq 0$, 
(iv) a finite momentum paired FFLO superfluid with $\Delta <\Delta _0$, $q\neq 0$, 
and $\delta n\neq 0$, and (v) a saddle point phase intervening between the local 
BCS and FFLO minima.}

\label{fig1}
\end{figure}

We now discuss separately the phase diagram in the situations where either
the field $\delta \mu$ or the population imbalance 
$\delta n=-\partial \Omega /\partial \delta \mu $ is kept fixed. To this end, we
trace the evolution of all available mean-field solutions with
increasing the dimensionless coupling constant $\eta =1/k_Fa$, where 
$k_F=(3\pi ^2n)^{1/3}$ is the non-interacting Fermi wave vector, and seek the
one with lowest energy (not the thermodynamic potential). To gain a physical
insight of the competing ground states, we show in Fig. 1 the landscape of
$\Omega$ at a selected set of parameters. At $\mathbf{q=0}$ there is a Sarma solution
situated between the trivial normal state at $\Delta=0$ and the local BCS minimum 
$\Delta _0$ and corresponding to a maxmium of $\Omega $ as a function of $\Delta $. 
On the other hand, for large enough field mismatch, a spatially modulated pairing
(known as FFLO phase) is driven with $\mathbf{q\cdot k}_F\sim \delta \mu $. 
This forms another local minimum in the landscape. Interestingly, a saddle point 
solution necessarily emerges in order to separate the local BCS and FFLO minima.

It is worth noting that not all the solutions are stable. As follows we mainly
focus on the stability against phase separation by the criterion 
$\partial \delta n/\partial \delta \mu >0$, which indicates the formation of
an inhomogeneous mixed state. Another stability criterion that the superfluid
density must be positive could also be readily examined \cite{yip}.

\begin{figure}
\begin{center}\includegraphics*[width=7cm]{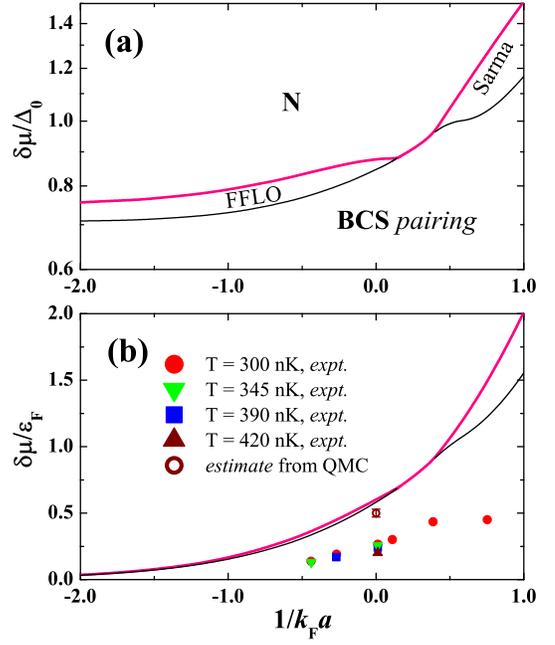}\end{center}

\caption{(color online). Phase diagram in the plane of interaction and
chemical potential difference.}

\label{fig2}
\end{figure}

\begin{figure}
\begin{center}\includegraphics*[width=6.5cm]{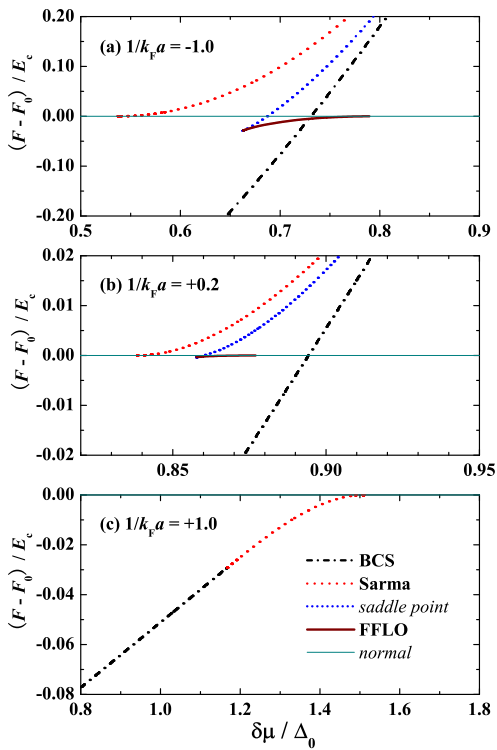}\end{center}

\caption{(color online). Comparison of free energies of available mean-field 
solutions at coupling constants as indicated, with the free energy of the normal 
gas $F_0$ being subtracted. $E_c={\cal N}(0)\Delta_{BCS}^2/2$ is the condensation 
energy for a symmetric Fermi gas, ${\cal N}(0)=mk_F/(2\pi ^2\hbar ^2)$ is the 
density of state at the Fermi surface and $\Delta _{BCS}=8\exp [\pi /(2k_Fa)-2]$.}

\label{fig3}
\end{figure}

{\it Fixed chemical potential difference}.---We present in Fig. 2a the interaction-field 
phase diagram, constructed by finding out the state with the lowest free energy 
$F=\Omega +\mu n$. The general structure of the phase diagram can be
understood by considering the BCS and BEC limits first. In the BCS limit
with infinitely small attraction, $\eta \rightarrow -\infty $, the kinetic
energy dominates and the Cooper pair formation is limited to the two
Fermi surface. For $\delta \mu <\delta \mu _1=1/\sqrt{2}\Delta _0$, the
ground state remains the BCS state. For $\delta \mu _1<\delta \mu <\delta
\mu _2\simeq 0.754\Delta _0$, the Fermi surfaces may be translationally
deformed, in order to increase the overlap for pairing. A FFLO state with spatially
varying order parameter is thereby more preferable. The transition from BCS
to FFLO states is of first order. Finally, for $\delta \mu >\delta \mu _2$,
the system translates continuously into a normal Fermi liquid phase. As an
example, for $\eta = -1$ we show in Fig. 3a the numerical comparison of
free energies of various competing states.

The ground state in the BEC limit of $\eta \rightarrow +\infty $ is also
known on physical grounds. Because of the strong attraction, all the spin
down fermions are likely to pair up with atoms in the other state, to form a
condensate of tightly bounded objects in real space. The distortion of Fermi
surfaces is prohibited, and then the leftover possibilities are the BCS
pairing and the Sarma state, as confirmed numerically in Fig. 3c. The latter
state, in this strong coupling limit, is a \emph{coherent} mixture of
condensate and a remaining Fermi sea of unpaired atoms. It is energetically
favorable only for $\delta \mu \simeq \epsilon _b$ as to create an unbound
fermion, where $\epsilon _b=\hbar ^2/2ma^2$ is the two-body binding energy.
For sufficiently large mismatch $\delta \mu \simeq \epsilon_b+2^{2/3}\epsilon _F$, 
the condensate disappears and the gas becomes completely polarized. 
Transitions among BCS, Sarma and normal phases are continuous.

The phase diagram in the two limits therefore are entirely different. Around
the BCS-BEC crossover one could image a qualitatively change. In particular,
the spatially varying FFLO and saddle point phases should cease to exist
with increasing the coupling. We find numerically (\textit{i.e.}, see Fig. 3b)
that for $0.15<\eta <0.40$ the system goes from BCS to the normal state,
without experiencing the FFLO nor the Sarma phase. Our mean-field finding is in
sharp contrast with a previous proposal in Ref. \cite{son}, where a direct
transition from FFLO to Sarma phase are anticipated. This anticipation is
another topological possibility to connect the two limits.

In Fig. 2b, by re-expressing $\delta \mu $ in terms of the non-interacting
Fermi energy, we compare our results of the critical $\delta \mu $ for
superfluid-to-normal transitions with the quantum Monte Carlo estimate \cite{carlson} 
and the recent experimental data on the critical Fermi energy difference 
$(\delta E_F/\epsilon _F)_c$ \cite{ketterle}. These differences are calculated 
assuming a non-interacting dispersion: 
$(\delta E_F/\epsilon _F)_c=[(1+(\delta n/n)_c)^{1/3}-(1-(\delta n/n)_c)^{1/3}]/2$, 
where $(\delta n/n)_c$ is the measured critical population imbalance (see, \textit{i.e.}, 
the Fig. 5 in Ref. \cite{ketterle}). The mean-field prediction is in good agreement with the 
Monte Carlo result, but is about two times larger than the measurement. This discrepancy 
should not be taken seriously since the mean-field theory is only qualitatively valid.
On the other hand, only in the weakly coupling BCS regime do the chemical potentials
equal the Fermi energies. Further, a quantitative comparison would require the 
consideration of the external trap. 

\begin{figure}
\begin{center}\includegraphics*[width=7cm]{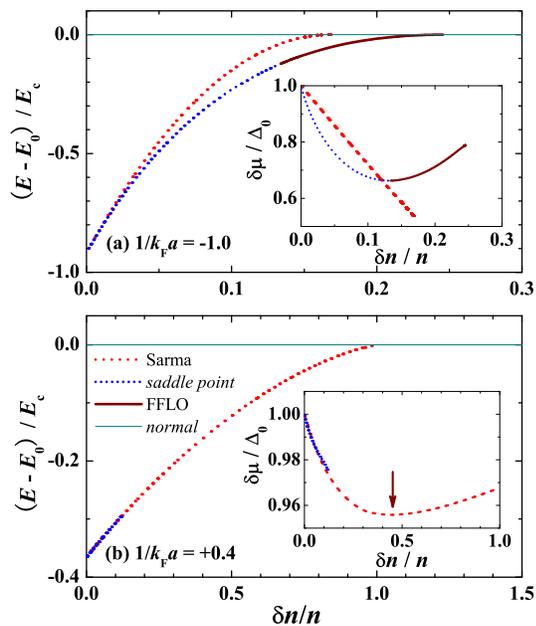}\end{center}

\caption{(color online). Comparison of energies of competing phases. Insets show 
the chemical potential difference as a function of polarization. The arrow 
in the inset of (b) indicates a position, above which the slope of the curve 
becomes positive.}

\label{fig4}
\end{figure}

\begin{figure}
\begin{center}\includegraphics*[width=7cm]{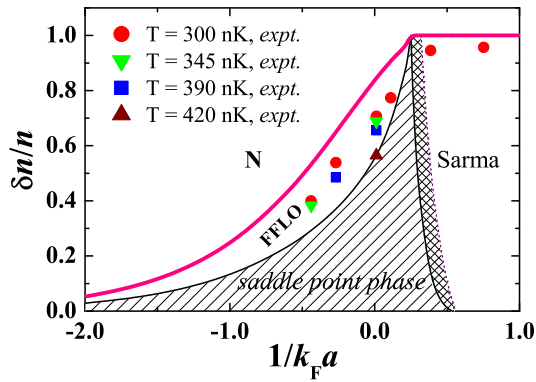}\end{center}

\caption{(color online). Interaction-porlarization phase diagram. The 
superfluid-to-normal transition boundary (thick line) is to be compared
with the experimental data (symbols). The shadowed region is unstable 
against phase separation.}

\label{fig5}
\end{figure}

{\it Fixed population imbalance}.---In this case, the phase diagram is
determined by minimizing $E=\Omega +\mu n+\delta \mu \delta n$. As shown in Fig.
4, now the spatially modulated saddle point phase and the FFLO phase are
energetically preferable if they exist. Therefore, on the BCS side, with
increasing imbalance the system goes from the saddle point state to the FFLO
state, and finally turns into a normal gas (Fig. 4a). As the interaction
strengths increase, the FFLO state disappears and the saddle point phase
also fades away, while the Sarma state starts to be supportive (Fig. 4b). In
the strong coupling BEC limit, the Sarma state becomes the only solution
left.

The above discussion yields a phase diagram in the plane of interaction and
polarization $\delta n/n$, as plotted in Fig. 5. It is topologically similar to
the diagram in the $\eta -\delta \mu $ plane, except that the BCS pairing
phase has now been replaced everywhere by the saddle point phase. However,
it is important to point out that the saddle point phase (shadow regions in
the figure), together with a sliver of the Sarma state, are intrinsically
unstable towards phase separation, since the slope of the plot of $\delta
\mu $ versus $\delta n$ for these phases is negative, as illustrated in the
insets of Fig. 4. This is exactly the \emph{precursor} for a spatially
inhomogeneous mixed phase \cite{bedaque}. Around the crossover, our
prediction for the appearance of the phase separation phase is consistent
with the experimental observations \cite{ketterle,hulet}.

In Fig. 5 we compare again the predicted boundary for the
superfluid-to-normal transition with the experimental findings of critical 
polarization $(\delta n/n)_c$ \cite{ketterle}. The agreement seems to be 
qualitatively good. We note, however, that the most intriguing FFLO state is 
not identified experimentally. The window for the FFLO state in our phase 
diagram is sizable, but it may shrink rapidly with increasing temperature 
and an external trap as in experiments.

We conclude by discussing the possible effects of quantum pair fluctuations 
beyond mean-field. Three remarks are in order concerning the $\eta -(\delta n/n)$ 
phase diagram. First, though within mean-field the BCS state is strictly confined 
to the horizontal axis ($\delta n=0$), the inclusion of the pair 
fluctuations may accommodate a finite population imbalance. As a result, a narrow window 
for a uniform BCS superfluid opens close to the axis of $\delta n=0$ inside 
the saddle point phase. Secondly, in our mean-field theory the phase boundary for 
the mixed phase is determined indirectly from an instability analysis. It can also be 
fixed following the way in Ref. \cite{bedaque}, \textit{i.e.}, by examining the 
energy of an incoherent mixture of some pure states. This alternative method 
requires considering of pair fluctuations on the strong coupling BEC side. Finally, 
so far we restrict our analysis to the free space. With a finite trap one may 
instead solve the mean-field Bogoliubov-de Gennes equations, or, resort to the 
local density approximation \cite{trap}. The latter approach is particularly useful 
in order to take into account the pair fluctuations in the presence of traps. 
Details of these issues on quantum fluctuations will be presented elsewhere.

We acknowledge simulating discussions with Prof. P. D. Drummond. This work
was supported by Australian Research Council Center of Excellence and the 
grant NSFC-10574080.


\begin{thebibliography}{10}
\bibitem{footnote}We assume the existence of a single $\mathbf{q}$ for FFLO states.
The general case of multiple $\mathbf{q}$ can also be examined \cite{rmp}.
\end{thebibliography}
\end{document}